\documentclass[prl,superscriptaddress,showpacs,twocolumn]{revtex4}
\usepackage{graphicx}
\usepackage{graphicx,amsfonts}
\usepackage{amsmath}
\usepackage{verbatim}

\usepackage{color}
\definecolor{darkblue}{rgb}{0,0,0.5}
\definecolor{darkred}{rgb}{0.5,0,0}
\usepackage[colorlinks=true,urlcolor=darkblue,citecolor=darkblue,linkcolor=darkred,hyperfootnotes=false]{hyperref}

\begin{document}

\newcommand{\atanh}
{\operatorname{atanh}}
\newcommand{\ArcTan}
{\operatorname{ArcTan}}
\newcommand{\ArcCoth}
{\operatorname{ArcCoth}}
\newcommand{\Erf}
{\operatorname{Erf}}
\newcommand{\Erfi}
{\operatorname{Erfi}}
\newcommand{\Ei}
{\operatorname{Ei}}

\newcommand{\qtot}{Q^{\tt tot}}
\newcommand{\qex}{Q^{\tt ex}}
\newcommand{\qhk}{Q^{\tt hk}}

\newcommand{\fqtot}{\mathcal{Q}^{\tt tot}}
\newcommand{\fqex}{\mathcal{Q}^{\tt ex}}
\newcommand{\fqhk}{\mathcal{Q}^{\tt hk}}

\newcommand{\rhoss}{\rho_{\tt SS}}
\newcommand{\Yhasa}{\mathcal{Y}}

\newcommand{\C}{\mathcal C}
\newcommand{\R}{\text{R}}
\newcommand{\hk}{\text{hk}}
\newcommand{\tot}{\text{tot}}
\newcommand{\Ss}{\mathcal S}
\newcommand{\ex}{\text{exp}}

\title{Unifying approach for fluctuation theorems from joint probability distributions}
\author{Reinaldo Garc{\'i}a-Garc{\'i}a}
\affiliation{Centro At\'omico Bariloche and Instituto Balseiro, 8400 S. C. de Bariloche, Argentina}
\author{Daniel Dom{\'i}nguez}
\affiliation{Centro At\'omico Bariloche and Instituto Balseiro, 8400 S. C. de Bariloche, Argentina}
\author{Vivien Lecomte}
\affiliation{Laboratoire de Probabilit\'es et Mod\`eles Al\'eatoires (CNRS UMR 7599), Universit\'e Pierre et Marie Curie - Paris VI – Universit\'e Paris-Diderot - Paris VII, 2 Place Jussieu, 75005 Paris, France}
\affiliation{DPMC-MaNEP, Universit\'e de  Gen\`eve, 24 Quai Ernest-Ansermet, 1211 Geneva 4, Switzerland}
\author{Alejandro B. Kolton}
\affiliation{Centro At\'omico Bariloche and Instituto Balseiro, 8400 S. C. de Bariloche, Argentina}

\begin{abstract}
Any decomposition of the total trajectory entropy production for
Markovian systems
has a joint probability distribution satisfying a generalized detailed
fluctuation theorem, when all the contributing terms are odd with
respect to time reversal. The expression of the result does not bring
into play dual probability distributions, hence easing potential
applications.
We show that several fluctuation theorems for perturbed
non-equilibrium steady states are unified and arise as particular cases 
of this general result. In particular, we show that
the joint probability distribution of the system and reservoir
trajectory entropies satisfy a detailed fluctuation theorem valid for
all times although each contribution does not do it separately.
\end{abstract}

\maketitle

The non-equilibrium stochastic thermodynamics of small systems has
attracted a lot of attention in the last years. From the experimental
side the development of techniques for microscopic manipulation has
allowed to study fluctuations in small systems of interest in Physics,
Chemistry and Biology~\cite{bustamante,ritort}. From the theoretical
side a group of exact relations known as fluctuation theorems
(FT)~\cite{Evans-Cohen-Morris,Gallavotti-Cohen,Kurchan,Lebowitz,Jarzynski,Crooks} has
shed light into the principles governing dissipation and fluctuations
in non-equilibrium phenomena, as in driven systems in contact with
thermal bath. Formally, the generality of the FTs can be attributed to
the way probability distribution functions of different observables
behave under time-reversal symmetry-breaking perturbations
(see~\cite{Kurchan2,Harris-Schuetz} for reviews on FT).

Oono and Paniconi~\cite{Oono-Paniconi} proposed a phenomenological
framework for a ``non-equilibrium steady-state (NESS) thermodynamics''
aimed at describing fluctuating systems subjected to an
external protocol. In this approach, the total exchanged heat during
a time interval $\tau$ by a system initially prepared in a NESS is
expressed as the sum of two contributions, $\qtot=\qex + \qhk$. The
``excess heat'' $\qex$ is, in average, associated with the energy
exchange during transitions between steady-states while the
``housekeeping heat'' $\qhk$ corresponds, in average, to the energy we
need to supply to maintain the system in a NESS. Hatano and
Sasa~\cite{Hatano-Sasa} introduced a formal framework for these
phenomenological ideas and derived a FT which extends the second law
of thermodynamics for transitions between NESS controlled by external
parameters $\sigma(t)$. The Hatano and Sasa FT is applicable to
trajectories $x(t)$ evolved with a Langevin or more generally a
Markovian dynamics starting from an initial condition sampled from a
NESS compatible with the initial values $\sigma(0)$ of the control
parameters. 

Under identical conditions different FTs were subsequently proposed
for NESS, involving the above decomposition of the exchanged heat. We
can distinguish between the so-called integral and detailed FTs for
systems initially prepared as described above. The integral
fluctuation theorems (IFT) are exact relations for the average over
histories of different stochastic trajectory functionals $W[x]$, such
as $\langle e^{-W}\rangle=1$. Examples are the Jarzynski FT for the total
work~\cite{Jarzynski}, the Hatano-Sasa
FT~\cite{Hatano-Sasa} and the Speck-Seifert FT~\cite{Seifert1}.
The so called detailed FT (DFT) are, on the
other hand, exact relations for the probability distributions
functions (PDF) of different observables $W$, such as
$P(W)/P^\R(-W)=e^{W}$ where $P^\R(W)$ corresponds to the time-reversed
protocol $\sigma^\R(t)=\sigma(\tau-t)$, and a NESS initial condition
compatible with $\sigma(\tau)$. Examples are given by the Crooks
relation~\cite{Crooks}, or Seifert fluctuation
theorem~\cite{Seifert2}. While observables satisfying a DFT trivially
satisfy an IFT, the opposite is not always true.  In many
recently formulated DFTs, a modified ``dual'' PDF $P^{\dagger R}(W)$ 
enters into play~\cite{Jarzynski1,Esposito},
which corresponds to trajectories generated by the adjoint dynamics
(defined below), in general different from the original dynamics of
the system. The presence of dual distributions is a strong limitation
to the experimental use of such DFT, or to theoretical applications
(\emph{e.g.} obtaining NESS generalization of fluctuation-dissipation
relations).
A central result of our work is that generalized DFTs can be
established {\it without relying on dual probabilities} by considering
\emph{joint} probability distributions for different complementary
observables, instead of single PDFs. The joint probability
distributions arise naturally from the above mentioned separation of
two contributions to the total heat exchanged in a
NESS. From this novel joint DFT all the known DFT and IFT follow in a
straightforward way.

Among the fluctuation theorems formulated for Markov dynamics, the
total trajectory entropy production $\Ss [x; \sigma]=\ln
(\mathcal{P}[x; \sigma]/\mathcal{P}[x^\R; \sigma^\R])$ plays a
fundamental role~\cite{Kurchan,Lebowitz,Seifert2}. Here
$\mathcal{P}[x; \sigma]$ ($\mathcal{P}[x^\R; \sigma^\R]$) denotes the
probability density of trajectory $x$ (time reversed trajectory $x^\R$)
in the forward (backward) protocol.  We include in $\mathcal{P}$ the
initial distribution of $x$.  We omit hereafter the
final time $\tau$ in all trajectory functionals, and use calligraphic
symbols to denote functionals, and normal symbols to denote their
values. 
It is straightforward to show that the total trajectory entropy production is
odd upon time-reversal: $\Ss[x^\R; \sigma^\R]=-\Ss[x; \sigma]$.  We show
that any decomposition of $\Ss$  in $M$ distinct
contributions, $\Ss [x; \sigma]=\sum_i^M \mathcal A_i[x;
\sigma]$, each of them being odd $\mathcal A_i[x^\R;
\sigma^\R]=-\mathcal A_i[x; \sigma]$, has a generating function
satisfying the symmetry
\begin{equation}
\left\langle e^{-\sum_i^M \lambda_i \mathcal A_i[x; \sigma]} \right\rangle =
\left\langle e^{-\sum_i^M (1-\lambda_i) \mathcal A_i[x; \sigma^\R]} \right\rangle_\R
\label{eq:symmetry}
\end{equation}
where $\lambda_i$ are arbitrary parameters and $\langle...\rangle_\R$
denotes average over trajectories in the reversed protocol. This
symmetry is equivalent to the following generalized DFT for the joint probability of the $\mathcal A_i[x;
\sigma]$'s:
\begin{equation}
\frac{P(A_1, A_2,..., A_M)}{P^\R(-A_1, -A_2,..., -A_M)}=e^{S}
\ \text{with}\ \  {S}=\sum_{i=1}^M A_i
\label{eq:joint}
\end{equation}
Note that the result involves no use of dual PDFs. To prove
\eqref{eq:symmetry} we start by noting that the average of any
observable $\mathcal{O}[x]$ over trajectories satisfies the relation
\begin{align}
 \langle\mathcal{O}[x;\sigma]\rangle
&= \int\mathcal{D}x\mathcal{P}[x^\R; \sigma^\R] \mathcal{O}[x;\sigma]e^{-\Ss [x^\R; \sigma^\R]} 
\nonumber \\
&= \int\mathcal{D}x\mathcal{P}[x; \sigma^\R] \mathcal{O}[x^\R;\sigma]e^{-\Ss [x; \sigma^\R]}
 \nonumber \\ 
&=\big\langle\mathcal{O}[x^\R;\sigma]e^{-\Ss [x; \sigma^\R]}\big\rangle_\R
\label{eq:identity}
\end{align}
where we have used $\Ss [x; \sigma]=-\Ss [x^\R; \sigma^\R]$
together with the change of variable $x\rightarrow x^\R$. Considering
$\Ss [x; \sigma]=\sum_i^M \mathcal A_i[x; \sigma]$ the proof comes around
$\langle e^{-\sum_i^M \lambda_i \mathcal A_i[x; \sigma]} \rangle =
\langle e^{-\sum_i^M \lambda_i \mathcal A_i[x^\R;\sigma]-\Ss [x; \sigma^\R]} \rangle_\R 
= \langle e^{-\sum_i^N (1-\lambda_i) \mathcal A_i[x; \sigma^\R]} \rangle_\R$.
Eq.\eqref{eq:joint} is proved in a similar way, or also follows
from~\eqref{eq:symmetry} since it is a symmetry for the generating
function of the joint distribution $P(A_1,...,A_M)$.

Before considering their particular application for explicit
decompositions of $\Ss$ we note that the symmetries~\eqref{eq:symmetry}
and~\eqref{eq:joint} are valid for all times $\tau$ for systems
prepared in any initial condition.  In particular, we see that the
{\it total trajectory entropy production FTs} $\langle e^{-\Ss }\rangle=1$
and $P(S)/P^\R(-S)=e^S$ hold without further assumption.

We will consider two generic frameworks where our result applies:
systems described (i) by Langevin dynamics, and (ii) by a Markov
dynamics on discrete configurations, exemplifying their parallel
features.
Let us first consider a particle driven out of equilibrium by a
constant force $f$ in a potential $U$, subjected to the Langevin
dynamics
\begin{equation}
\dot{x} =-\partial_x U(x;\alpha(t))+f+\xi
\label{eq:langevin}
\end{equation}
where $\alpha(t)$ represents a set of control parameters, and $\xi(t)$
the Langevin noise, $\langle\xi(t) \rangle =0$,
$\langle\xi(t)\xi(t')\rangle=2T\delta(t-t')$, modelling the
interaction of the system with a thermal bath at temperature $T$.  We
consider for simplicity a single degree of freedom $x$, but our results
are easily generalized \emph{e.g.} in larger dimensions and/or with
more particles.  For a stochastic trajectory generated
by~\eqref{eq:langevin} we define the total exchanged heat as~\cite{Sekimoto}
\begin{equation}\label{totalheat}
 \beta \fqtot[x;\sigma]=-\beta\int_{0}^{\tau}dt\;\dot{x} 
 \big[\partial_x U (x;\alpha)-f \big]
\end{equation}
The total exchanged heat in a trajectory can be split
as $\fqtot=\fqhk+\fqex$~\cite{Oono-Paniconi}.
Defining $\phi(x;\sigma)=-\ln \rhoss(x,\sigma)$ from the steady-state probability
density $\rhoss(x,\sigma)$ at fixed values of $\sigma=(\alpha,f)$ Hatano and Sasa~\cite{Hatano-Sasa}
proposed
\begin{equation}\label{housekeeping functional}
 \beta \fqhk[x;\sigma]=
\int_{0}^{\tau}dt\;\dot{x}
\big[
{\partial_x\phi}(x;\sigma)-\beta\left({\partial_x U}(x;\alpha)-f\right)
\big]
\end{equation}
for the housekeeping heat and
\begin{equation}\label{excessheat}
 \beta \fqex[x;\sigma]=-\int_{0}^{\tau}dt\;\dot{x}{\partial_x\phi(x;\sigma)}
\end{equation}
for the the excess heat. The Hatano-Sasa functional~\cite{Hatano-Sasa} $\Yhasa[x;\sigma]$ is then defined as
\begin{equation}\label{HSfunctional}
\Yhasa[x;\sigma]\equiv\int_{0}^{\tau}dt\;\dot{\sigma}{\partial_\sigma\phi(x;\sigma)}
= \beta \fqex[x;\sigma]+\Delta\phi(x;\sigma)
\end{equation}
where
$\Delta\phi(x;\sigma)=\phi(x(\tau);\sigma(\tau))-\phi(x(0);\sigma(0))$
is a time boundary term. 

We now assume that the system is initially prepared in a NESS
compatible with $\sigma(0)$. With the previous definitions of
Eqs.(\ref{totalheat}),(\ref{housekeeping
  functional}),(\ref{excessheat}) and (\ref{HSfunctional}) it is known
that the total trajectory entropy production $\Ss $ can be
decomposed as the sum of two contributions, in two different
ways~\cite{Seifert1}
\begin{equation}
\label{particularentropyprod}
\Ss =\Yhasa + \beta \fqhk=\Delta\phi+\beta \fqtot
\end{equation}

Let us now show that similar decompositions also exist for Markov
dynamics: we consider discrete configurations $\{\C\}$ with
transition rates $W(\C\to\C';\sigma)$ between configurations. They
depend on $\sigma$, an external control parameter which may vary in
time. The probability density at time $t$ obeys the Markov dynamics
$\partial_t P(\C,t)=\sum_{\C'} W(\C'\to\C;\sigma(t)) P(\C',t) -
r(\C;\sigma(t))P(\C,t)$ where $r(\C;\sigma)=\sum_{\C'}
W(\C\to\C';\sigma)$ is the escape rate from configuration $\C$. An
history of the system is described by the succession of configuration
$(\C_0,\ldots,\C_K)$ visited by the system, $\C_k$ being visited
between $t_k$ and $t_{k+1}$.  Starting from initial distribution
$P_{\text{i}}(\C,\sigma)$, the probability of an history is
$
  \mathcal{P}[\C;\sigma]= 
  e^{-\int_0^\tau dt\, r(\C(t);\sigma(t))}
  \prod_{k=1}^K W(\C_{k-1}\to\C_k,\sigma_{t_k})
  P_{\text{i}}(\C(0),\sigma(0)) 
$ 
meaning that the mean value of an history-dependent observable $\mathcal O$
is given by
$
  \langle\mathcal O\rangle
= \sum_{K\geq 0}
 \sum_{\C_{0}\ldots \C_{K}}
\!\!\int_0^t dt_K 
\ldots \int_0^{t_2} dt_1
 \mathcal O[\C,\sigma]
 \mathcal{P}[\C,\sigma]
$. 
We obtain that the total trajectory entropy production $\Ss [\C; \sigma]=\ln
(\mathcal{P}[\C; \sigma]/\mathcal{P}[\C^\R; \sigma^\R])$ has a first
decomposition $\Ss =\Delta\phi+\beta \fqtot$ with
$
  \Delta\phi=\log \frac{ P_{\text{i}}(\C(0),\sigma(0))}{ P_{\text{i}}(\C(\tau),\sigma(\tau))} 
$ and
\begin{align}
  \beta \fqtot= \sum_{k=1}^K \log \frac{ W(\C_{k-1}\to\C_k,\sigma_{t_k})}{W(\C_{k}\to\C_{k-1},\sigma_{t_k})} 
\label{eq:def_beta-Qhk-Markov}
\end{align}
Although there is no natural definition of $\beta$ we
keep the notation $\beta \fqtot$ to exemplify the parallel with
Langevin dynamics. 

Turning to the second decomposition, let's now assume that the initial distribution
is steady-state: $P_{\text{i}}=\rhoss=e^{-\phi}$.
One directly checks that the Hatano-Sasa functional
$\Yhasa[\C,\sigma]=\int_0^\tau dt\:\dot\sigma  \partial_{\sigma}\phi$
writes
\begin{align}
 \Yhasa = \big[\phi(\C,\sigma)\big]_0^\tau - 
 \sum_{k=1}^K\big[\phi(\C_k,\sigma_{t_{k}}) - \phi(\C_{k-1},\sigma_{t_k})\big]
\label{eq:HSfunctional_decomposed}
\end{align}
Besides, defining the house-keeping work as
\begin{align}
 \beta \mathcal Q^\hk[\C,\sigma] 
=& 
 \sum_{k=1}^K \log\frac{W(\C_{k-1}\to\C_k,\sigma_{t_k})}{W(\C_k\to\C_{k-1},\sigma_{t_k})}
\nonumber \\ & + 
 \sum_{k=1}^K\phi(\C_k,\sigma_{t_{k}}) - \phi(\C_{k-1},\sigma_{t_k})
\end{align}
we check that the decomposition $\Ss =\Yhasa + \beta \fqhk$
holds~\cite{us}.  The parallel between Markov and Langevin frameworks
also appears by specializing to rates $W(k\to k\pm 1)=e^{-\frac
  \beta 2(V_{k\pm 1}-V_k)}$ of jumping on a 1d lattice from site $k$ to
$k\pm 1$ in a tilted potential $V_k=U_k-kf$: in the continuum limit,
one recovers the Langevin observables~\cite{us}.

In the first decomposition, $\Yhasa$ can be identified with
the so-called non-adiabatic contribution (since it vanishes for
quasistatic protocols) to the trajectory entropy $\Ss _{\rm
na} \equiv \Yhasa$ while $\beta \fqhk$ (which is continuously
produced in the steady-state) can be identified with the so-called
adiabatic part $\Ss _{\rm a} \equiv \beta \fqhk$
~\cite{Esposito}. On the other hand, in the second decomposition
of $\Ss $, $\Delta \phi$ can be identified with the system
contribution $\Ss _{s}\equiv \Delta \phi$ while $\beta
\fqtot$ can be identified with the reservoir contribution
$\Ss _{r}\equiv \beta \fqtot$ to the total trajectory
entropy production.

The entropy decompositions of~\eqref{particularentropyprod} satisfy the conditions for the
application of the identities~\eqref{eq:symmetry} or~\eqref{eq:joint}
since each term is odd with respect to time reversal in both
decompositions. We can thus write DFTs (valid for all times
$\tau$) for the joint probabilities as
\begin{align}
\label{joint1}
\frac{ P(Y, \beta \qhk)}{ P^\R(-Y, -\beta \qhk)} &= e^{Y+\beta \qhk} \\
\label{joint3} \frac{ P(\Delta\phi, \beta \qtot)}{
P^\R(-\Delta\phi, -\beta \qtot)} &= e^{\Delta\phi+\beta \qtot}
\end{align}
with~\eqref{joint3} valid for any initial distribution with
$
  \Delta\phi=\log \frac{ P_{\text{i}}(\C(0),\sigma(0))}{ P_{\text{i}}(\C(\tau),\sigma(\tau))} 
$
while~\eqref{joint1} requires starting from the NESS.
The corresponding IFTs write
\begin{eqnarray}
\label{joint2} \langle e^{-\lambda\Yhasa-\kappa\beta\fqhk}\rangle
&=& \langle
e^{-(1-\lambda)\Yhasa_\R-(1-\kappa)\beta\fqhk_\R}\rangle_\R \\
\label{joint4} \langle e^{-\lambda\Delta\phi-\kappa\beta
\fqtot}\rangle &=& \langle
e^{-(1-\lambda)\Delta\phi_\R-(1-\kappa)\beta \fqtot_\R}\rangle_\R
\end{eqnarray}
Here $\mathcal{X}_\R$ denotes here $\mathcal{X}[x;\sigma^\R]$,
$\mathcal{X}$ representing $\Yhasa$, $\fqhk$, $\fqtot$ or
$\Delta\phi$. From~(\ref{joint1}) and (\ref{joint3}) we
have, in terms of $S_s$, $S_r$, $S_a$ and $S_{\rm na}$, that
$P(S_s,S_{r})/P^\R(-S_s,-S_{r})=e^{S_s+ S_{r}}$ and
$P(S_a,S_{na})/P^\R(-S_a,-S_{na})=e^{S_a+S_{na}}$. It is worth noting
that these relations do not involve dual PDFs, and thus they can be
tested for a physical system with a given dynamics. We also note that
while one can show that $S_a$ and $S_{na}$ satisfy each one separately
a DFT by using dual PDFs~\cite{Esposito}, $S_s$ and $S_r$ satisfy a
joint DFT although they do not satisfy separately a DFT.

Let us now derive from an unified view the known FTs. As an
intermediate step, it is useful to define  a ``dual'' trajectory
weight $\mathcal{P}^\dagger[x]$  as
~\cite{Jarzynski1,Esposito}
\begin{equation}
\label{eq:pdagger} \mathcal{P}^\dagger[x;\sigma] =
\mathcal{P}[x;\sigma] e^{ -\beta\fqhk[x;\sigma]}.
\end{equation}
For Markov dynamics the dual
probability $\mathcal P^\dagger$ corresponds to the dynamics in the
so-called dual rates $W^\dagger(\C\to\C',\sigma) \equiv
e^{-[\phi(\C',\sigma)-\phi(\C,\sigma)]} W(\C'\to\C,\sigma) $ which
share the same steady state as the original dynamics.
In the case of the Langevin dynamics of Eq.~\eqref{eq:langevin}, it corresponds
to trajectories  generated by the equation~
$\dot{x} =   -\partial_x U^\dagger(x;\alpha(t))+f^\dagger+\xi$
with $U^\dagger = 2\phi/\beta-U$, $f^\dagger=-f$. This equation
also has the same steady state as the original one. In both cases,
let us stress that the dual dynamics corresponds to 
trajectories in a different physical system.
We now follow~\eqref{eq:pdagger} to write the
dual joint PDF related to~\eqref{joint1} as
\begin{equation}\label{dual}
 P^\dagger(Y, \beta \qhk)=P(Y, -\beta \qhk)e^{\beta \qhk}
\end{equation}
which is normalized.
Integrating this relation over $Y$, we first obtain the DFT~\cite{Esposito}
$P(\beta\qhk)=e^{\beta \qhk }P^\dagger(-\beta\qhk)$, and hence the IFT
$\langle e^{-\beta \qhk } \rangle=1$~\cite{Seifert1}.
Using now successively~\eqref{joint1} and~\eqref{dual}
\begin{align}
  P(Y)&= e^{Y}\int d(\beta \qhk)\;e^{\beta \qhk}P^\R(-Y,-\beta \qhk)
\nonumber\\
  \label{step2hatano}
  &=e^{Y}\int d(\beta \qhk)\;P^{\dagger R}(-Y,\beta \qhk)
\end{align}
we see that the DFT $P(Y)=e^{Y}P^{\dagger R}(-Y)$~\cite{Jarzynski1}
holds.  This implies the corresponding
IFT $\langle e^{-\Yhasa}\rangle =1$~\cite{Hatano-Sasa}
(also derived from setting
$\lambda=1$, $\kappa=0$ in~\eqref{joint2} and using the
Speck-Seifert IFT).

As an illustration of our approach, let us show how joint FTs provide
insights on the experimental error in the evaluation of entropy
productions. Consider an experiment where the steady state can be
evaluated for different values of the control parameter $\sigma$,
\emph{e.g.} microspheres optically driven in a
liquid~\cite{Trepagnier}. Having in hand an experimental evaluation
$\phi_\ex$ of $\phi$, we write
\begin{equation}
  \Ss=\Yhasa_\ex+\delta\Yhasa+\beta\fqhk
  \label{eq:SwithYYexp}
\end{equation}
where $\Yhasa_\ex[x;\sigma]=\int_0^\tau dt
\dot\sigma\partial_\sigma\phi_\ex$ and
$\delta\Yhasa=\Yhasa-\Yhasa_\ex$ is the difference between exact and
experimental Hatano-Sasa functionals. Starting from the NESS
associated to $\sigma(0)$, each of the terms in~\eqref{eq:SwithYYexp}
is odd upon time-reversal, and we can use Eq.~(\ref{eq:joint}) for $M=3$,
which yields the DFT 
$$
P(Y_\ex,\delta Y,\beta\qhk) = P^{\R}(-Y_\ex,-\delta Y,-\beta\qhk) e^{Y_\ex+\delta Y+\beta\qhk}.
$$ 
Using~\eqref{dual}, this gives
$
P(Y_\ex,\delta Y) = P^{\dag\R}(-Y_\ex,-\delta Y)e^{Y_\ex+\delta Y}
$ 
and hence also the IFT
\begin{equation}
  \big\langle e^{-\Yhasa_\ex}  \big\rangle = 
  \big\langle e^{-\delta\Yhasa_\R}  \big\rangle_\R^\dag, 
\end{equation}
which allows to estimate the difference between the 
IFT $\langle e^{-\Yhasa}\rangle =1$ and the experimentally
obtained $\langle e^{-\Yhasa_\ex}  \rangle$.

 As a second example let
us consider the situation in which the system, initially prepared
in a NESS, has a variation of its parameters
$\sigma_i(t)=\sigma_i^0+\delta\sigma_i(t)$ in such a way that
$|\frac{\delta\sigma_i(t)}{\sigma_i^0}|\ll1$, with
$\sigma_i^0=\sigma_i(0)$ and $\delta\sigma_i(0)=0$. In this
context a modified Fluctuation-Dissipation Theorem has been
recently derived in Ref. \cite{Parrondo} which relates dissipation
under small perturbations around a NESS, with fluctuations in the
corresponding steady state. 
One can expand the exponentials in~\eqref{joint2} up to second
order in $\delta\sigma$ and then in powers of $\lambda$ and
$\kappa$. To second order in $\lambda$ and order zero in $\kappa$,
 we arrive to (see \cite{us} for details),
\begin{equation}\label{correlation}
 \langle\mathcal{B}_{ij}(0,\tau)\rangle_0=
\langle\mathcal{B}_{ij}(\tau,0)
e^{-\beta\int_0^{\tau}dt\dot{x}(t)v_s(x(t);\sigma^0)}\rangle_0,
\end{equation}
where $\mathcal{B}_{ij}(t,t')=\frac{\partial\phi(x(t);\sigma^0)}{\partial\sigma_i}
\frac{\partial\phi(x(t');\sigma^0)}{\partial\sigma_j}$
and $v_s=\beta^{-1}\partial_x\phi-(\partial_x U - f)$ corresponds to the average
 velocity in the NESS associated to $\sigma$. 
For systems with Boltzmann-Gibbs steady state,
the obtained result reduces to the symmetry
$\mathcal{C}_{ij}(\tau)= \mathcal{C}_{ij}(-\tau)$, with
$\mathcal{C}_{ij}(\tau)=\langle\mathcal{B}_{ij}(0,\tau)\rangle_0$. 
Eq.(\ref{correlation}) can  also be
derived from Eq.(\ref{step2hatano}) in Ref.\cite{Jarzynski1}.
However, the use of the joint PDF can lead us to obtain further
new results. Let us introduce a weighted correlation function as
\begin{equation}\label{weightedcorrelation}
 \mathcal{C}_{ij}^W(\tau,0)=
 \frac{\langle\mathcal{B}_{ij}(0,\tau)
 e^{-\frac{\beta}{2}\int_0^{\tau}dt\dot{x}(t)v_s(x(t);\sigma^0)}\rangle_0}
{\langle e^{-\frac{\beta}{2}\int_0^{\tau}dt\dot{x}(t)v_s(x(t);\sigma^0)}\rangle_0}
\end{equation}
This correlation function carries explicit information about the
lack of detailed balance and reduces to the usual one when
the system is able to equilibrate. Using
Eq.(\ref{joint2}) with $\kappa=\frac{1}{2}$ and
repeating the same procedure we have done in order to obtain
Eq.(\ref{correlation}) we arrive to the result
$\mathcal{C}_{ij}^W(\tau,0)=\mathcal{C}_{ij}^W(0,\tau)$, which is
completely symmetric and reduces to the known result for
equilibrium dynamics when detailed balance holds.

In conclusion, the identities~\eqref{eq:symmetry} and~\eqref{eq:joint}
and their immediate consequences for Markovian systems are the main
message of this work. Equations~(\ref{eq:symmetry})
or~\eqref{eq:joint} indeed contain, as particular cases, several known
FTs such as the ones previously derived by Hatano and Sasa
\cite{Hatano-Sasa}, Speck and Seifert~\cite{Seifert1},
Chernyak {\it et al}.~\cite{Jarzynski1} and Esposito and Van den Broek
\cite{Esposito}. In addition, an exact DFT, valid for all times
$\tau$, holds for the joint distribution of the reservoir and system
entropy contributions to the total trajectory entropy production, although each
contribution does not do it separately, as given by
Eq.(\ref{joint3}). Also a similar DFT holds for the joint distribution
of the adiabatic and nonadiabatic entropy contributions to the total
trajectory entropy, as given by Eq.(\ref{joint1}). It is worth to
mention that for the type of NESS discussed here, $M=2$ decompositions
of the total trajectory entropy production are obtained,
Eq.(\ref{particularentropyprod}), and thus two-variable joint PDFs are
all that is needed for the corresponding DFTs. We have shown
and example with $M=3$ for handling experimental errors in the
Hatano-Sasa FT. In any case, in the light
of~\eqref{eq:symmetry} and~\eqref{eq:joint}, obtaining an adequate
minimal $M$-decomposition of the total trajectory entropy production
constitutes the cornerstone towards the derivation of generalized FTs
for non-equilibrium systems.

\begin{acknowledgments}
This work was supported by CNEA, CONICET (PIP11220090100051), ANPCYT (PICT2007886), 
V.L. was supported in part by the Swiss NSF under MaNEP and Division II and 
thanks CNEA for hospitality.
\end{acknowledgments}


\begin{thebibliography}{100}

\bibitem{bustamante}
C. Bustamante, J. Liphardt, F. Ritort,
Physics Today, {\bf 58} 43 (2005).

\bibitem{ritort}
F. Ritort,
Advances in Chemical Physics \textbf{137}, 31 (2008) Ed. Wiley $\&$ Sons. arXiv:0705.0455v1.

\bibitem{Evans-Cohen-Morris}
Denis J. Evans, E. G. D. Cohen and G. P. Morriss,  Phys. Rev. Lett. {\bf 71}, 2401 (1993).

\bibitem{Gallavotti-Cohen}
G. Gallavotti and E. G. D. Cohen, Phys. Rev. Lett. {\bf 74}, 2694 (1995).

\bibitem{Kurchan}
J. Kurchan, J. Phys. A: Math. Gen. {\bf 31} 3719 (1998).

\bibitem{Lebowitz}
J. L. Lebowitz and H. Spohn, J. Stat. Phys. {\bf 95} 333 (1999).

\bibitem{Jarzynski}
C. Jarzynski, Phys. Rev. Lett. {\bf 78}, 2690 (1997); C. Jarzynski, Phys. Rev. E {\bf 56}, 5018 (1997).

\bibitem{Crooks}
G. E. Crooks, J. Stat. Phys. {\bf 90}, 1481 (1998); G. E. Crooks, Phys. Rev. E {\bf 61}, 2361 (2000).


\bibitem{Kurchan2}
J. Kurchan, J. Stat. Mech. (2007) P07005

\bibitem{Harris-Schuetz}
R. J. Harris and G. M. Sch\"utz, J. Stat. Mech. P07020

\bibitem{Oono-Paniconi}
Y. Oono and M. Paniconi, Prog. Theor. Phys. Suppl. {\bf 130}, 29 (1998)

\bibitem{Hatano-Sasa}
Takahiro Hatano and Shin-ichi Sasa, Phys. Rev. Lett. {\bf 86}, 3463 (2001).

\bibitem{Seifert1}
T. Speck and U. Seifert, J. Phys. A: Math. Gen. {\bf 38}, 581 (2005)

\bibitem{Seifert2}
U. Seifert, Phys. Rev. Lett. {\bf 95} 040602 (2005)

\bibitem{Jarzynski1}
V. Y Chernyak, M. Chertkov and C. Jarzynski, J. Stat. Mech. P08001 (2006).

\bibitem{Esposito}
M. Esposito and C. Van den Broeck, Phys. Rev. Lett. \textbf{104}, 090601 (2010)

\bibitem{Sekimoto}
K. Sekimoto, Prog. Theor. Phys. Suppl. \textbf{130}, 17 (1998)

\bibitem{us}
R. Garc{\'i}a-Garc{\'i}a, V. Lecomte, A. B. Kolton and D. Dom{\'i}nguez, in preparation.


\bibitem{Trepagnier}
E. H. Trepagnier, C. Jarzynski, F. Ritort, G. E. Crooks, C. J. Bustamante and J. Liphardt,
PNAS \textbf{101} 15038 (2004).

\bibitem{Parrondo}
J. Prost, J.-F. Joanny and J. M. R. Parrondo, Phys. Rev. Lett. {\bf 103} 090601 (2009)




\end{thebibliography}
\end{document}